\documentclass[aps,pre,preprint,onecolumn,footinbib,superscriptaddress]{revtex4-1}
\pdfoutput=1
\usepackage{amsmath,graphicx,amssymb}
\usepackage{subfigure,float}
\usepackage{epsfig}
\usepackage{color,url}
\usepackage{verbatim}
\usepackage{natbib}
\usepackage{bm}

\renewcommand{\vec}[1]{\bm{#1}}
\newcommand{\etal}{\textit{et al.}}
\newcommand{\Ceps}{C_\varepsilon}
\newcommand{\Cinf}{C_{\varepsilon,\infty}}

\newcommand{\vep}{\varepsilon}
\newcommand{\RL}{R_L}
\newcommand{\Rl}{R_{\lambda}}

\newcommand{\cdiminf}{C_{\vep,\infty}}

\newcommand{\beq}{\begin{equation}}
\newcommand{\eeq}{\end{equation}}

\begin{document}

\title{Energy transfer and dissipation in forced isotropic 
turbulence}

 \author{W.~D. McComb}
 \affiliation{
 SUPA, School of Physics and Astronomy,
 University of Edinburgh, James Clerk Maxwell Building, The King's Buildings, Edinburgh EH9 3JZ, UK}
 \author{A. Berera}
 \affiliation{
 SUPA, School of Physics and Astronomy,
 University of Edinburgh, James Clerk Maxwell Building, The King's Buildings, Edinburgh EH9 3JZ, UK}
 \author{S.~R. Yoffe}
 \affiliation{
 SUPA, Department of Physics, 
 University of Strathclyde, John Anderson Building, 107 Rottenrow East, Glasgow G4 0NG, UK}
 \author{M.~F. Linkmann}
 \affiliation{
 SUPA, School of Physics and Astronomy,
 University of Edinburgh, James Clerk Maxwell Building, The King's Buildings, Edinburgh EH9 3JZ, UK}

\begin{abstract}
A model for the Reynolds number dependence of the dimensionless
dissipation rate $C_{\varepsilon}$ was derived from the dimensionless
K\'{a}rm\'{a}n-Howarth equation, resulting in
$C_{\varepsilon}=C_{\varepsilon, \infty} + C/R_L + O(1/R_L^2)$, where
$R_L$ is the integral scale Reynolds number. The coefficients $C$ and
$C_{\varepsilon,\infty}$ arise from asymptotic expansions of the
dimensionless second- and third-order  structure functions. This
theoretical work was supplemented by direct numerical simulations (DNSs)
of forced isotropic turbulence for integral scale  Reynolds numbers up
to $R_L=5875$ ($\Rl=435$), which were used to establish that the decay
of dimensionless dissipation with increasing Reynolds number took the
form of a power law $R_L^n$ with exponent value $n = -1.000\pm 0.009$, and that this
decay of $C_{\varepsilon}$ was actually due to the increase in the
Taylor surrogate $U^3/L$.  The model equation was fitted to data from
the DNS which resulted in the value $C=18.9\pm 1.3$ and in an asymptotic
value  for $C_\varepsilon$ in the infinite Reynolds number limit of
$C_{\varepsilon,\infty} = 0.468 \pm 0.006$.
\end{abstract}

\maketitle

\section{Introduction}

In recent years there has been much interest in the fundamentals of
turbulent dissipation, as characterized by the mean dissipation rate
\beq
\label{eq:eps_defn}
\vep=\frac{\nu_0}{2}\sum_{\alpha, \beta=1}^3 \left \langle \left ( \frac{\partial u_{\alpha}}{\partial x_{\beta}} 
+ \frac{\partial u_{\beta}}{\partial x_{\alpha}} \right )^2 \right  \rangle \ ,
\eeq
where $\nu_0$ is the kinematic viscosity, $u_{\alpha}\equiv u_{\alpha}(\vec{x},t)$ is one component of the 
velocity field $\vec{u}$, while angle brackets denote an ensemble average. 
For isotropic turbulence, \eqref{eq:eps_defn} reduces to
\beq
\label{eq:eps_defn_iso}
\vep=\nu_0\sum_{\alpha, \beta=1}^3 \left \langle \left ( \frac{\partial u_{\alpha}}{\partial x_{\beta}}  \right )^2 \right  \rangle \ .
\eeq
This interest has centered on the approximate
expression for the dissipation rate $\vep$, which was given by Taylor 
in 1935 \cite{Taylor35} as
\beq
\vep = \Ceps U^3/L,
\label{Taylor-diss}
\eeq
where $U$ is the root-mean-square velocity and $L$ is the integral
scale. Many workers in the field refer to Eq.~(\ref{Taylor-diss})
as the \emph{Taylor dissipation surrogate}. However, others re-arrange it
to define the coefficient $\Ceps$ as the nondimensional dissipation
rate; thus,
\beq
\Ceps = \frac{\vep}{U^3/L}.
\label{dim-diss}
\eeq
In 1953 Batchelor \cite{Batchelor71} (we refer to the first edition of this work) 
presented evidence to
suggest that the coefficient $\Ceps$ tended to a constant value with
increasing Reynolds number. 
In 1984 Sreenivasan \cite{Sreenivasan84}
showed that in grid turbulence $\Ceps$ became constant for
Taylor-Reynolds numbers greater than about $50$.
He also found a $1/\Rl$-dependence at low $\Rl$ and, since at low $\Rl$
the Taylor-Reynolds number and the integral scale Reynolds number are 
proportional, Sreenivasan's paper had already in effect presented empirical 
evidence for $1/R_L$ scaling at low $R_L$. We discuss this further,
in relation to our present work, in Section \ref{sec:numerics}.    
Later, in 1998, Sreenivasan
presented a survey of investigations of both forced and decaying
turbulence \cite{Sreenivasan98}, using direct numerical simulation
(DNS), which established the now characteristic curve of $\Ceps$ plotted
against the Taylor-Reynolds number $\Rl$ (e.g.~see our Fig.~\ref{fig:Ceps_Rl}).
More recently, the comprehensive review of dissipation rate scaling by Vassilicos 
\cite{Vassilicos15} has summarized the evidence for $1/R_L$ scaling of $\Ceps$. 

In his 1968 lecture notes \cite{Saffman68}, Saffman made two comments
about the expression that we have given here as Eq.~(\ref{Taylor-diss}). 
These were as follows: ``This result is fundamental to an
understanding of turbulence and yet still lacks theoretical support''
and  ``the possibility that $A$ (i.e.~our $\Ceps$) depends weakly
on the Reynolds number can by no means be completely discounted.'' More
than 40 yr on, the question implicit in his second comment has
been comprehensively answered by the survey papers of Sreenivasan
\cite{Sreenivasan84,Sreenivasan98}, along with a great deal of
subsequent work by others, some of which we have cited here. However,
while some theoretical work has indicated an inverse proportionality
between $\Ceps$ and Reynolds number, this has been limited to low 
Reynolds numbers \cite{Sreenivasan84} or based on a
mean-field approximation \cite{Lohse94} or restricted to providing an
upper bound \cite{Doering02}.  Hence, his first comment is still valid
today; and this lack of theoretical support remains an
impediment to the development of turbulence phenomenology and hence
turbulence theory.

In this article we present two pieces of work. These are as follows. 

First we develop a theoretical model of the relationship
between the dimensionless dissipation rate and the integral scale
Reynolds number. We start from the driven Navier-Stokes equation in
wavenumber space and specify the nature of the input term to the energy
balance equation in wavenumber space. Then we Fourier transform this in
order to derive the energy balance in scale space, that is, the
K\'{a}rm\'{a}n-Howarth equation with forcing. This provides a basis for the
application of our general  theory for forced isotropic turbulence to the
specific case of our DNS driven by negative damping. It also gives a
basis for a later consideration of the universality of our conclusions.

Second, we present the data obtained from DNS for a range of integral
scale Reynolds numbers up to $R_L =5875$. These results are used to
elucidate some aspects of the phenomenon and then to test our
theoretical model.

We begin with a short review of the relevant literature. 

\section{Some results from both numerical and experimental investigations}

Unless otherwise  stated, the cited DNSs used the standard
pseudospectral method  simulating isotropic turbulence in cubic boxes of
length  $L_{box}=2 \pi$. We report  on results for forced isotropic
turbulence only. Some of the numerical results mentioned below are shown
in Fig.~\ref{fig:Ceps_Rl} alongside our data.
 
Jim\'{e}nez \etal~\cite{Jimenez93} attained Taylor-scale Reynolds
numbers up  to $\Rl=170$, with their highest $\Rl$ simulation extending
to $0.3\tau$, where  $\tau$ denotes the large eddy turnover time. In
view of the short execution time this  simulation might still be in a
transient state. They achieved dealiasing by a combination of
random grid shifts and spherical truncation. The system was forced  by using
negative viscosity for wavenumbers $k \leqslant 2.5$ maintaining 
$k_{max}\eta$ and hence $\varepsilon$ constant,  where $\eta$ denotes
the Kolmogorov dissipation scale. The authors reported an asymptotic
value for the dimensionless dissipation rate $\Cinf \simeq 0.7$. The
statistics were calculated from five to ten  realizations for a short
execution time. That is, given the sample rate and the  run time, the
realizations would have been strongly correlated. Regarding resolution
requirements,  the authors point out that $k_{max}\eta = 1$ is the
absolute minimum while  $k_{max}\eta = 2$ is desirable. 

In the work of Wang \etal~\cite{Wang96} 
the forcing was implemented by maintaining the kinetic energy in the two lowest 
wavenumber shells constant with an energy spectrum following $k^{-5/3}$. 
The measured asymptote $\Cinf$ lay in the region 
$0.42 \leqslant \Cinf \leqslant 0.49$. Using the same method without 
dealiasing, Cao \etal~\cite{Cao99} focused mainly on the statistics of the
pressure field, but data is provided in their Table 1 from which $\Ceps$ can be 
calculated. The initial condition was similar to our DNS as 
$E(k,0) \sim k^4\exp(k/k_0)^2$, with $k_0 \simeq 5$ and the system evolved 
for ten large eddy turnover times before measurements were taken.

Yeung and Zhou \cite{Yeung97} presented time-averaged results from
simulations using a partially dealiased code with stochastic forcing, covering
a $\Rl$ range of $38 \leq \Rl \leq 240$ for about four large-eddy turnover 
times. The resolution was relatively high as all runs satisfied 
$k_{max}\eta \geq 1.5$. 

A partially dealiased code with stochastic forcing was also used by Donzis 
\etal~\cite{Donzis05}, who simulated flows with Taylor-scale Reynolds number up 
to $\Rl=390$. The data points for $\Ceps$ at different $\Rl$ were fitted to 
the expression $\Ceps= A(1+\sqrt{1+(B/\Rl)^2})$, with 
$A\simeq 0.2$ and $B \simeq 92$, leading to an asymptote $\Cinf \simeq 0.4$.
We discuss this expression for $\Ceps$ in more detail in 
Sec.~\ref{sec:asym_exp}.

The investigation by Bos \etal~\cite{Bos07} reported results from DNS, 
Large Eddy Simulation (LES) and 
Eddy-Damped, Quasi-Normal Markovian closure (EDQNM) calculations for 
Reynolds numbers up to $\Rl=100$ for DNS and $\Rl=2000$ 
for EDQNM. The authors tested different initial conditions such as 
Gaussian-shaped initial energy spectra 
and the von K\'{a}rm\'{a}n spectrum and found no dependence on the choice of
initial spectrum once the system had reached a stationary state. However, 
the transient to a steady state was found to be 
shorter for a von K\'{a}rm\'{a}n spectrum than for Gaussian-shaped initial
 spectra. They measured $\Cinf \simeq 0.5$ for the asymptote of the 
dimensionless dissipation rate. 

Variations of the initial conditions were also studied by Goto and
Vassilicos \cite{Goto09},  mainly by altering the low wave number
behavior and  the peak wave number of the initial spectra. The results
for $\Ceps$ show a dependence on the different low wave number forms of
the initial spectra.  In contrast, the location of the peak of the
initial spectrum had no significant  influence on $\Ceps$. What is
interpreted as a dependence on the form of the initial spectra could
actually be due to differences in the forcing method. The system is
kept statistically stationary by fixing the magnitude of the  velocity
field modes for wave numbers smaller than the peak wave number of the 
initial spectra, which in some cases leads to a very large forcing
range. The low wave number form of the initial spectrum is thus
maintained during the evolution  of the velocity field, such that it is
no longer purely a feature of the initial  condition but rather a
permanent feature imposed by the forcing scheme. The observed dependence
of $\Ceps$ on the choice of initial energy spectrum could  therefore be
due to differences in the forcing spectrum instead.

Kaneda \etal~\cite{Kaneda03} conducted the largest DNS of forced
isotropic  turbulence so far on grids of up to $4096^3$ collocation
points reaching  $\Rl = 1201$ in single precision and $\Rl=732$ in
double precision, both at minimum resolution of $k_{max}\eta =1$. The
system was maintained  statistically stationary by using negative
viscosity for wave numbers  $k\leqslant 2.5$ in order to keep the total
energy constant. Data were collected from single realizations only,
resulting in an asymptotic  value for $\Ceps$ in the range $0.4
\leqslant \Cinf \leqslant 0.5$. The largest  $\Rl$ simulation was only
carried out for a short time; thus, this run might  still be transient. 

The most recent high resolution DNS results for the dimensionless
dissipation  rate were presented by Yeung \etal~\cite{Yeung12}.  Four
simulations spanning a Taylor-scale Reynolds number range of  $140
\leqslant \Rl \leqslant 1000$ on $2048^3$ and $4096^3$ collocation
points were carried out, at resolutions between  $1.3 \leqslant
k_{max}\eta \leqslant 11.2$, resulting  in $0.449 \leqslant \Ceps
\leqslant 0.470$. Due to the computational  cost incurred by simulations
of this size, the execution time in steady state was  relatively short
and the simulation corresponding to $\Rl=1000$ was stopped  after
$3.59\tau$. During the steady state, 20 snapshots were taken to populate
the ensemble, so samples were taken every  $0.18\tau$. Thus the ensemble
consisted of realizations that are statistically correlated.  The
authors noted that a longer run time would be preferable, but argued
that since intense fluctuations in $\vep$ are relatively short lived,
 ensemble averaging over snapshots close in time will still improve
statistics. 

In contrast to the various pseudospectral DNSs of incompressible
turbulent flows  cited here, Pearson \etal~\cite{Pearson04a} used a
sixth-order finite  difference scheme with large-scale
$\delta(t)$-correlated forcing for DNS of slightly  compressible flows,
leading to $\Ceps \simeq 0.5$.   

Having summarized numerical results on the topic we now briefly turn to 
experimental results. Pearson \etal~\cite{Pearson02}  measured $\Ceps
\simeq 0.5$ for a  number of shear flows. Different flow types were
investigated by Burattini \etal~\cite{Burattini05}, and Mazellier
\etal~\cite{Mazellier08} studied turbulence in a wind tunnel generated
from a variety of  different grid geometries including fractal grids. In
the fractal case they  found a significantly lower asymptote for
$\Ceps$, namely, $\Cinf \simeq 0.065$.  However, we should note that
turbulence generated in this way differs in other quite profound ways
from conventional grid turbulence.

In all, we find that the asymptotic value $\Cinf \simeq 0.5$ is a
well-established  numerical result which is broadly in agreement with
experimental work.

\section{A model for the dependence of dimensionless dissipation 
on Reynolds number}

The use of external random forcing with the Navier-Stokes equations
(NSEs) was pioneered in the development of statistical theories in the
late 1950s. This work was very much influenced by problems in
statistical  physics, such as Brownian motion, and the emphasis was on
choosing forces which could lead to turbulence that was characteristic
of the NSE, rather than the forcing. For this reason we begin with a
spectral formulation. However, it is also convenient in that it allows us to
make a connection with our DNS, which employs the usual pseudospectral
method. We obtain the energy balance in wavenumber space (the Lin
equation), and then Fourier transform this to obtain the energy balance
in scale space. The result is, of course, fully equivalent to the
K\'{a}rm\'{a}n-Howarth equation with forcing, as derived entirely by more
conventional means; see Chap.~4 in the book \cite{McComb14a}. In
obtaining our theoretical model for the dimensionless dissipation rate, we
introduce the dimensionless K\'{a}rm\'{a}n-Howarth equation and make asymptotic
expansions of the structure functions in inverse powers of the integral
scale Reynolds number. We first consider the idealized problem of
isotropic turbulence with $\delta-$function forcing in wave number and then
apply the analysis to the finite forcing used in the DNS.

\subsection{Energy balance and the nature of the forcing}

In Fourier space, the incompressible NSEs may be
written as:
\begin{align}
\label{NSE}
(\partial_t + \nu_0 k^2)\vec{u}(\vec{k},t) & = 
i\vec{k}P(\vec{k},t)
+\int_{\mathbb{R}^3} d \vec{j} \ (i\vec{k} \cdot \vec{u}(\vec{j},t)) \vec{u}(\vec{k}-\vec{j},t)
+ \vec{f}(\vec{k},t) \ , \nonumber \\
i\vec{k} \cdot \vec{u}(\vec{k},t)=0 \ ,
\end{align}
where $\vec{u}(\vec{k},t)$ denotes the three-dimensional Fourier transform 
of the velocity field $\vec{u}(\vec{x},t)$, $P(\vec{k},t)$ the Fourier transform of 
the pressure field, $\nu_0$ the kinematic viscosity, and $\vec{f}(\vec{k},t)$
the Fourier transform of the stirring force $\vec{f}(\vec{x},t)$. In
order to avoid introducing unwanted correlations into the problem, the
stirring forces must be highly uncorrelated in time. For this reason,
they are normally taken to have delta-function autocorrelations in time;
see \cite{Kraichnan59a,Edwards64,McComb90a,McComb14a}. In other
statistical problems, this input is often referred to as \emph{white noise}.

The energy balance in wavenumber space (the Lin equation) can readily be
derived from the above NSE (see \cite{McComb14a}) to obtain the well
known form
\begin{equation}
 \frac{\partial E(k,t)}{\partial t} = T(k,t) - 2\nu_0 k^2E(k,t) + W(k,t) \ ,
\label{eq:Lin}
\end{equation}
where $E(k,t)$ and $T(k,t)$ are the energy 
and transfer spectra, respectively, and 
\beq
W(k,t) = 4\pi k^2 \langle \vec{u}(-\vec{k},t)\cdot\vec{f}(\vec{k},t) \rangle 
\eeq 
is the work spectrum of the stirring force. For conciseness we do not
explicitly show the time dependence from now on.

In order to avoid introducing a dependence on the forcing in wave number
space, it was argued by Edwards in 1965 that the forcing spectrum could
take the form of a $\delta$-function at the origin. In a modern notation
\cite{McComb14a}, this may be written as
\beq
\label{eq:delta}
W(k) = \varepsilon_W \delta (k) ,
\eeq
thus introducing the injection rate $\varepsilon_W$ which, in more
general terms, is defined by
\beq
\label{eq:epsW}
\varepsilon_W = \int^\infty_0 W(k)\ dk \ .
\eeq
At this point we note that $W(k)$ is integrable, which follows from 
the well posed nature of the problem, as both $\vec{f}$ and $\vec{u}$ should 
be square-integrable in order to ensure that the total energy remains finite 
(and to ensure the existence of the respective Fourier transforms).
 
An alternative to the use of stirring forces exists in the form of
negative damping at low wave numbers. This was introduced to theoretical
work in 1966 by Herring \cite{Herring66} and to numerical simulation by
Machiels in 1997 \cite{Machiels97a}. It is now quite widely used and, as
in several of the investigations cited herein, it was used in our
present DNS. In this method, the Fourier transform of the force is given by
\begin{align}
 \vec{f}(\vec{k},t) &=
      (\varepsilon_W/2 E_f) \vec{u}(\vec{k},t) \quad
\text{for} \quad  0 < \lvert\vec{k}\rvert < k_f , \nonumber \\
  &= 0   \quad \textrm{otherwise},
\label{forcing}
\end{align}
$E_f$ being the total energy contained in the forcing
band. This ensures that the energy injection rate is $\varepsilon_W =
\textrm{constant}$. The highest forced wavenumber, $k_f$, is usually taken 
to be small. This form of energy input was used in our numerical simulations, 
as discussed in Section \ref{sec:numerics}.

\subsection{The K\'{a}rm\'{a}n-Howarth equation for forced turbulence}

Now we obtain the equivalent form of the K\'{a}rm\'{a}n-Howarth equation
(KHE), by Fourier transformation of the Lin equation \cite{McComb14b} as
\beq
 \label{gen-khe}
 -\frac{3}{2}\frac{\partial U^2}{\partial t} +\frac{3}{4}\frac{\partial 
 S_2(r)}{\partial t} = -\frac{1}{4r^4} \frac{\partial}{\partial r} \Big( 
 r^4 S_3(r) \Big) 
 + \frac{3\nu_0}{2r^4} \frac{\partial}{\partial
r}\left( r^4 \frac{\partial S_2(r)}{\partial r} \right)
 - I(r) \ , 
\eeq
where the longitudinal structure functions are defined as
\begin{equation}
 S_n(r) = \left\langle \big([\vec{u}(\vec{x}+\vec{r}) - 
 \vec{u}(\vec{x})]\cdot\vec{\hat{r}}\big)^n \right\rangle \ ,
\end{equation}
and the input $I(r)$ is given in terms of $W(k)$ by
\begin{align}
 I(r) 
 \label{eq:I_expr}
      &= 3 \int_0^\infty dk\ W(k)\ \left[ \frac{\sin{kr} - 
      kr\cos{kr}}{(kr)^3} \right] \ ,
\end{align}
where the convergence of this integral is a consequence of the integrability of 
$W(k)$ ensured by the well posed nature of this problem as stated below \eqref{eq:epsW} 
in the previous section.
Here $I(r)$ is interpreted as the total energy injected into all scales $> 
r$.
Note that we may make the connection between $W(k)$ and the injection rate 
for the numerical simulations by
\beq
 \label{eq:I0}
 I(0) = \int^\infty_0 dk\, W(k) = \vep_W \ ,
\eeq
where the energy injection rate $\vep_W$ is as specified for the DNS by \eqref{forcing}.

It is also helpful to introduce the energy decay rate $\vep_D=-(3/2)\partial U^2/\partial t$,
and with some rearrangement \eqref{gen-khe} may be written as 
\beq
 \label{eq:gen-khe_epsD}
 \vep_D= - \frac{3}{4}\frac{\partial S_2(r)}{\partial t}  
-\frac{1}{4r^4} \frac{\partial}{\partial r} \Big(r^4 S_3(r) \Big)
 + \frac{3\nu_0}{2r^4} \frac{\partial}{\partial
r}\left( r^4 \frac{\partial S_2(r)}{\partial r} \right)
 - I(r) \ .
\eeq

At this stage we have a general form of the KHE, but it does not contain
the dissipation rate as such (irrespective of how the KHE is derived).
As it is the dissipation rate which interests us, we may introduce it to
the KHE by a simple identity. This can be derived by integration of the
Lin equation \eqref{eq:Lin} with respect to wavenumber. Hence, one
obtains for the energy balance of isotropic turbulence
\beq
-\vep_D= 0 - \vep + \vep_W \ ,
\label{eq:E_balance} 
\eeq
as $\int dk \ T(k) =0$, by conservation of energy; see \cite{McComb14a}. 

For freely decaying turbulence, where $\vep_W=0$, this relation becomes $\vep_D=\vep$.
Hence, the rate of change of the total energy is due to dissipation only, 
as expected. 

For forced turbulence which has reached a stationary state there is no change in the  
total energy. That is, $\vep_D=0$, and the dissipation rate must equal the rate of energy 
input; hence, $\vep=\vep_W$.

If we substitute \eqref{eq:E_balance} into \eqref{eq:gen-khe_epsD} we obtain the 
most general form of the KHE 
\beq
 \label{eq:gen-khe_epsW}
 \vep-\vep_W= - \frac{3}{4}\frac{\partial S_2(r)}{\partial t}  
-\frac{1}{4r^4} \frac{\partial}{\partial r} \Big(r^4 S_3(r) \Big)
 + \frac{3\nu_0}{2r^4} \frac{\partial}{\partial
r}\left( r^4 \frac{\partial S_2(r)}{\partial r} \right)
 - I(r) \ ,
\eeq 
which can be applied either to forced and/or to decaying turbulence by
setting the appropriate  terms to zero.   

That is, if we were to apply \eqref{eq:gen-khe_epsW} to freely-decaying turbulence, we
would set the input term $I(r)$ equal to zero, to give
\begin{equation}
 \vep_D \equiv \vep = -\frac{3}{4}\frac{\partial 
 S_2}{\partial t} -\frac{1}{4r^4} \frac{\partial}{\partial r} \Big( r^4 S_3 
 \Big) + \frac{3\nu_0}{2r^4} \frac{\partial}{\partial r}\left( r^4 
 \frac{\partial S_2}{\partial r} \right) \ ,
\end{equation}
which is the form of the KHE familiar in the literature (e.g.~see \cite{McComb14a} 
or \cite{Monin75}). 

Here we are considering forced turbulence which has reached a
stationary state. So we must set the left-hand side of \eqref{eq:gen-khe_epsW}
and any time-derivatives that appear in this equation, such as $\partial
S_2/\partial t$, to zero. Whereupon (\ref{eq:gen-khe_epsW}) reduces 
(with some rearrangement) to the appropriate KHE for forced turbulence,
\begin{equation}
 \label{eq:fKHE}
 I(r) =  -\frac{1}{4r^4} \frac{\partial}{\partial r} \Big( r^4 S_3(r) \Big) 
 + \frac{3\nu_0}{2r^4} \frac{\partial}{\partial r}\left( r^4 \frac{\partial 
 S_2(r)}{\partial r} \right) \ .
\end{equation}
After an integration with respect to $r$, this equation is further rearranged to take 
the form 
\begin{equation}
\label{eq:S_work}
 S_3(r) = -\frac{4}{r^4} \int_0^r dy\ y^4 I(y) + 6\nu_0\frac{\partial 
 S_2}{\partial r} \ ,
\end{equation}
where $I(r)$ contains \emph{all} the information of the forcing and is calculated
directly from the work spectrum.
If we take the limit $r\to0$ in Eq.~\eqref{eq:I_expr}, and
invoke stationarity, then for
small scales we obtain $\lim_{r \to 0} I(r) = \varepsilon_W = \varepsilon$, and so
recover the Kolmogorov form of the KHE \cite{Kolmogorov41b} from \eqref{eq:fKHE}
\begin{equation}
 \vep=\vep_W =  -\frac{1}{4r^4} \frac{\partial}{\partial r} \Big( r^4 S_3(r) \Big) 
 + \frac{3\nu_0}{2r^4} \frac{\partial}{\partial r}\left( r^4 \frac{\partial 
 S_2(r)}{\partial r} \right) \ ,
\end{equation}
for small scales. Alternatively, at the other extreme, with the Edwards $\delta$-function
forcing \eqref{eq:delta}, this relationship holds for all scales.
However, a middle ground can be found if, instead of taking a limit, we restrict 
our attention to scales below the forcing scale, where the energy input to scale $r$ 
is independent of the details of the forcing.


\subsection{Dimensionless K\'{a}rm\'{a}n-Howarth equation for stationary turbulence}

Returning to our form of the forced KHE, Eq.~\eqref{eq:fKHE}, we now introduce the 
dimensionless structure functions $h_n(\rho)$
which are given by
\begin{equation}
 S_n(r) = U^n h_n(\rho) \ ,
\end{equation}
where $\rho = r/L$. Substitution of these into \eqref{eq:fKHE} leads to 
\beq
    I(\rho) = -\frac{1}{4\rho^4} \frac{\partial}{\partial \rho} \Big(\rho^4 h_3(\rho) \Big)\frac{U^3}{L}
            + \frac{\nu_0U^2}{L^2} \frac{3}{2\rho^4} \frac{\partial}{\partial\rho}\left(
              \rho^4 \frac{\partial h_2(\rho)}{\partial \rho} \right) \ .
\eeq
Then, with some re-arrangement, \eqref{eq:fKHE} takes the dimensionless form
\beq
  \label{eq:dimless_KHE_genforce}
    I(\rho)\frac{L}{U^3} = -\frac{1}{4\rho^4} \frac{\partial}{\partial \rho} \Big(\rho^4 h_3(\rho) \Big)
            + \frac{1}{\RL} \frac{3}{2\rho^4} \frac{\partial}{\partial\rho}\left(
              \rho^4 \frac{\partial h_2(\rho)}{\partial \rho} \right) \ ,
\eeq
with $R_L = UL/\nu_0$ the Reynolds number based on the integral scale.
For conciseness we introduce coefficients $A_3$ and $A_2$: 
\beq
 \label{eq:A3}
 A_3(\rho) = -\frac{1}{4\rho^4} \frac{\partial}{\partial \rho} \Big(
\rho^4 h_3(\rho) \Big) \ , 
\eeq
and
\beq
 \label{eq:A2}
A_2(\rho) = \frac{3}{2\rho^4} \frac{\partial}{\partial
 \rho}\left( \rho^4 \frac{\partial h_2(\rho)}{\partial \rho} \right) \ ,
\eeq
Equation \eqref{eq:dimless_KHE_genforce} expressed in terms of $A_2$ and 
$A_3$ then becomes
\begin{equation}
 I(\rho) \frac{L}{U^3} = A_3(\rho) + \frac{A_2(\rho)}{R_L} \ .
\label{dim-input}
\end{equation}

The input $I(\rho)$ may be expressed in terms of an amplitude $\varepsilon_W$ and a 
dimensionless shape function $\phi(\rho)$ thus:
\begin{equation}
 \label{eq:phi}
 I(\rho) = \varepsilon_W \phi(\rho) \ ,
\end{equation}
where $\phi(\rho)$ contains all of the scale-dependent information and, as 
required by Eq.~\eqref{eq:I0},
satisfies $\phi(0) = 1$. Using the shape function $\phi$, Eq.~\eqref{dim-input}
reads
\begin{equation}
\phi(\rho) \frac{\vep_WL}{U^3}= A_3(\rho) + \frac{A_2(\rho)}{R_L} \ ,
\label{dim-input_shapeftc}
\end{equation}
where the left-hand side already looks similar in structure to the dimensionless
dissipation rate $\Ceps=\vep L/U^3$.

Now let us consider the dimensionless KHE for the case of 
constant forcing at the small scales, where $\phi(\rho)=1$; 
hence, $I(\rho) =  \vep_W$. Equation 
\eqref{dim-input} becomes
\begin{equation}
 \frac{\vep_W L}{U^3} = A_3(\rho) + \frac{A_2(\rho)}{R_L} \ ,
\end{equation}
from which, since $\vep = \vep_W$ from stationarity, and using Eq.~\eqref{dim-diss}, we 
have
\begin{equation}
 \Ceps = \frac{\vep_W L}{U^3}= A_3(\rho) + \frac{A_2(\rho)}{R_L} \ .
 \label{pre-DA-model}
\end{equation}
This simple scaling analysis has extracted the integral scale as the 
relevant lengthscale, and $R_L$ as the appropriate Reynolds number, for 
studying the behavior of $\Ceps$, but it is not unique. If we had used 
different scales, the coefficients $A_2$ and $A_3$ would also be different.
This particular scaling was advocated by Batchelor 
\cite{Batchelor71}, despite which it has become common practice to study 
$\Ceps = \Ceps(R_\lambda)$, as shown in Fig. \ref{fig:Ceps_Rl}.

From the well-known phenomenology associated with Kolmogorov's
inertial-range theories \cite{Kolmogorov41b}, as the Reynolds number
tends to infinity, we know that we must have $A_2/\RL \to 0$ 
and $A_3\to \Cinf = \mbox{constant}$. 

Equation \eqref{pre-DA-model} can also be rewritten as
\begin{equation}
 \label{eq:pre_DA_model_rearr}
 \varepsilon = A_3(\rho)\frac{U^3}{L} + 
 A_2(\rho)\frac{\nu_0U^2}{L^2} \ .
\end{equation}
The first term on the right-hand side is essentially the Taylor surrogate, while the 
second term is a viscous correction. It has been demonstrated \cite{McComb10b} 
that, for the case of decaying turbulence, the surrogate $U^3/L$ represents 
the maximum inertial transfer flux, $\varepsilon_T$, more accurately 
than the dissipation rate. Here $\vep_T$ is given by the maximum of the 
transport power $\Pi_{max}$, 
\beq
\vep_T= \Pi_{max}=\int_{k^*}^{\infty} dk \ T(k) \ ,
\eeq
where $k^*$ denotes the single zero crossing of the transfer spectrum; 
for further details, see p.~88 in \cite{McComb14a}. 
The same is shown later for forced turbulence in Fig.~\ref{fig:Taylor_surrogate}, since the 
input rate (and hence $\varepsilon$) is kept constant. Thus, the forced KHE  
expresses the equivalence of the rates at which energy is transferred and 
dissipated (or injected) as $\nu_0 \to 0$. For finite viscosity, there is a 
contribution to the dissipation rate which has not passed through the 
cascade. In terms of our re-arranged model equation, we may write 
\eqref{eq:pre_DA_model_rearr}
\begin{equation}
 \varepsilon = \Cinf\frac{U^3}{L} + \nu_0 \frac{A_2(\rho)U^2}{L^2} \to 
 \varepsilon_T \qquad\text{as}\qquad \nu_0 \to 0 \ ,
\end{equation}
where, from Eq.~\eqref{eq:A3}, the asymptotic value denoted by $\Cinf$ is given by the expression
\begin{equation}
 \Cinf = \lim_{\nu_0 \to 0} A_3(\rho)  
= -\lim_{\nu_0 \to
0} \frac{1}{4\rho^4}\frac{\partial}{\partial \rho}
\left(\rho^4 h_3(\rho) \right)   \ .
\label{eq:onset_ff}
\end{equation}

At this point we note that taking the limit $\nu_0 \to 0$ in \eqref{eq:onset_ff} 
corresponds to the onset of Kolmogorov's four-fifths law and that
therefore the existence of the constant $\Cinf$ 
corresponds to the same physical situation as 
the four-fifths law 
\cite{Batchelor53,Tennekes72,Pope00}. 

\subsection{Asymptotic expansion of the structure functions in inverse powers of $\RL$}
\label{sec:asym_exp}

In order to examine the dependence of the dimensionless dissipation rate
on $\RL$ in detail, it is convenient to go back to the form of energy
balance [i.e.~\eqref{eq:dimless_KHE_genforce}] that we had before we
introduced the coefficients $A_2$ and $A_3$. Restricting our attention
to scales smaller than the energy injection scale, we have
$I(\rho)=\vep_W =\vep$, hence the dimensionless KHE
\eqref{eq:dimless_KHE_genforce} reads
\beq
  \label{eq:dimless_KHE_expl}
    \Ceps = -\frac{1}{4\rho^4} \frac{\partial}{\partial \rho} \Big(\rho^4 h_3(\rho) \Big)
            + \frac{1}{\RL} \frac{3}{2\rho^4} \frac{\partial}{\partial\rho}\left(
              \rho^4 \frac{\partial h_2(\rho)}{\partial \rho} \right) \ .
\eeq
This expression already suggests a dependence of $\Ceps$ on $\RL$. However, the structure functions, and hence
their dimensionless counterparts $h_2(\rho)$ and $h_3(\rho)$, also depend on Reynolds number. In order to treat 
their Reynolds-number dependence, we consider asymptotic expansions in inverse powers of $\RL$. 

We note that for large $\RL$ the term with the highest derivative in \eqref{eq:dimless_KHE_expl} 
is multiplied by the small parameter $\RL^{-1}$, hence we are faced with 
a singular perturbation problem \cite{Lundgren02}. Therefore, we consider outer asymptotic expansions of the structure 
functions in negative powers of $\RL$, a technique applied to singular perturbation problems
(see e.g.~\cite{Wasow65}, Chap.~X). We study here only the outer expansions as we have 
rescaled the KHE with respect to the integral scale $L$.

The outer expansions of the dimensionless structure functions in powers of $\RL^{-1}$ are
   \beq
     \label{eq:exp_h2}
     h_2(\rho)=h_2^{(0)}(\rho) + \frac{1}{\RL}h_2^{(1)}(\rho) + O\left(\frac{1}{\RL^2}\right) \ , 
   \eeq
   and 
   \beq
     \label{eq:exp_h3}
     h_3(\rho)=h_3^{(0)}(\rho) + \frac{1}{\RL}h_3^{(1)}(\rho) + O\left(\frac{1}{\RL^2}\right) \ . 
   \eeq
Substituting the expansions \eqref{eq:exp_h2} and \eqref{eq:exp_h3} into \eqref{eq:dimless_KHE_expl}
we obtain up to first order in $\RL^{-1}$
   \beq
     \label{eq:exp_KHE}
    \Ceps = -\frac{1}{4\rho^4} \frac{\partial}{\partial \rho} \Big(\rho^4 h^{(0)}_3(\rho) \Big)
            + \frac{1}{\RL} \left [ \frac{3}{2\rho^4} \frac{\partial}{\partial\rho}\left( 
              \rho^4 \frac{\partial h^{(0)}_2(\rho)}{\partial \rho} \right)
            -\frac{1}{4\rho^4} \frac{\partial}{\partial \rho} \Big(\rho^4 h^{(1)}_3(\rho) \Big) \right ] 
            + O\left(\frac{1}{\RL^2}\right)\ ,     
   \eeq   
   where the terms $h_2^{(0)}$, $h_3^{(0)}$, and $h_3^{(1)}$ do not depend on $\RL$.
 We can write this in terms of the coefficient $\Cinf$ and a new coefficient $C$, 
 both of which are constant with respect to $R_L$. Thus, 
  \beq
   \label{eq:flux}
   \cdiminf = -\frac{1}{4\rho^4} \frac{\partial}{\partial \rho} \Big(\rho^4 h^{(0)}_3(\rho) \Big) 
  \eeq
  and 
  \beq
   \label{eq:visc_corr}
   C = \frac{3}{2\rho^4} \frac{\partial}{\partial\rho}\left(
              \rho^4 \frac{\partial h^{(0)}_2(\rho)}{\partial \rho} \right)
            -\frac{1}{4\rho^4} \frac{\partial}{\partial \rho} \Big(\rho^4 h^{(1)}_3(\rho) \Big) \ ,
  \eeq
  where both coefficients are \emph{a priori} scale dependent (i.e.~dependent on a length scale), 
while $\Ceps$ is not. Hence, the scale dependencies of the different terms in the model 
equation must cancel each other. 
In fact, since $\Cinf$ is a constant with respect to $\rho$ by the four-fifths law, the scale 
dependence between the two terms on the right-hand side of \eqref{eq:visc_corr} must 
cancel out. This leads us to the model equation  
  \beq
   \label{eq:DA_model}
    \Ceps=\Cinf + \frac{C}{R_L} \ ,
  \eeq
  where $\Cinf$ and $C$ are constants with respect to $R_L$ \emph{and} $\rho$. 

In order to compare with results plotted against Taylor-Reynolds number $\Rl$, we 
substitute the relation 
\beq
R_L=\Ceps R_{\lambda}^2/15 \ ,
\eeq 
into 
\eqref{eq:DA_model} and solve for $\Ceps$. This leads to an expression for the 
dependence of $\Ceps$ on $R_\lambda$, 
\beq
\Ceps(\Rl) = A \big(1 + \sqrt{1 + (B/R_\lambda)^2} \big) \ , 
\label{eq:Ceps_Rl}
\eeq
where $A$ and $B$ are constants with respect to $\Rl$. We note that this
particular step was first taken by Doering and Foias \cite{Doering02},
who derived an expression similar to \eqref{eq:DA_model} as an upper
bound on the dependence of $\Ceps$ on $\RL$.

\section{Numerical Simulations}
\label{sec:numerics}

We used the standard pseudospectral method with full dealiasing for our DNS;
further details can be found in Ref.~\cite{McComb14b}.
The initial conditions were Gaussian-distributed random velocity fields with a
prescribed energy spectrum of the form
\beq
E(k,0) \sim k^4\exp(k/k_0)^2 \ ,
\eeq
with $k_0 \simeq 5$.
The system was forced at the large scales by negative damping as in \eqref{forcing} with 
$k_f \leqslant 2.5$. This method has also been used in other investigations
\cite{Jimenez93,Yamazaki02,Kaneda03,Kaneda06}, albeit not necessarily
such that $\vep_W$ is maintained constant.

For each Reynolds number studied, we used the same initial spectrum and
input rate $\varepsilon_W$. The only initial condition changed was the
value assigned to the (kinematic) viscosity $\nu_0$. Note that increasing the Reynolds
number by decreasing $\nu_0$, at constant $\vep_W$ is the same as taking the infinite 
Reynolds number limit. 

Measurements were taken after the simulations had reached a
stationary state, determined by the {\em mean} total energy becoming
constant: for a discussion of this criterion, see \cite{McComb01a}, and
in particular Fig. 3 of that reference.
The velocity field was sampled every half a large-eddy
turnover time, $\tau = L/U$, where $L$ denotes the average integral scale and
$U$ the rms velocity.
The ensemble populated with these sampled realizations was used,
in conjunction with the usual shell averaging, to
calculate statistics. Simulations were run using lattices of size
$128^3$ up to $2048^3$, with corresponding Taylor-Reynolds
numbers ranging from $R_\lambda = 41.8$ up to $435.2$.
All simulations were sufficiently resolved at the small scales, that is the
maximum wavenumber satisfied $k_\text{max}\eta
 \geqslant 1.30$ for all runs except one which satisfied
$k_\text{max}\eta \geqslant 1.01$, where $\eta$ is the Kolmogorov dissipation
lengthscale. Large-scale resolution has only relatively recently
received attention in the literature. The integral scale, $L$, was found to lie between
$0.23 L_{\text{box}}$ and $0.17 L_{\text{box}}$; that is, the largest scales
of the flow are smaller than a quarter of the simulation box size.
Details of the simulations are summarized in Table \ref{tbl:simulations}.

\begin{table}[tb!]
 \begin{center}
  \begin{tabular}{ccllllllll}
  $ R_L $ & $R_\lambda$ & $\nu_0$ & $N$ & $\varepsilon$ & $\sigma$ & $U$ & 
  $L/L_\text{box}$ & $k_\text{max}\eta$ & $t_{ss}/\tau$ \\
  \hline
  \hline
81.5 &  41.8  & 0.01    & 512  & 0.097 & 0.010 & 0.581 & 0.22 & 9.57 & 12.61\\
83.7 &  42.5  & 0.01    & 128  & 0.094 & 0.015 & 0.581 & 0.23 & 2.34 & 12.06\\
88.2 &  44.0  & 0.009   & 128  & 0.096 & 0.009 & 0.587 & 0.22 & 2.15 & 12.74\\
101.4 &  48.0  & 0.008   & 128  & 0.096 & 0.013 & 0.586 & 0.22 & 1.96 & 12.72\\
105.7 &  49.6  & 0.007   & 128  & 0.098 & 0.011 & 0.579 & 0.20 & 1.77 & 13.82\\
146.5 &  60.8  & 0.005   & 512  & 0.098 & 0.009 & 0.589 & 0.20 & 5.68 & 14.09\\
158.6 &  64.2  & 0.005   & 128  & 0.099 & 0.011 & 0.607 & 0.21 & 1.37 & 13.80\\
287.8 &  89.4  & 0.0025  & 512  & 0.101 & 0.006 & 0.605 & 0.19 & 3.35 & 15.20\\
360.1 &  101.3 & 0.002   & 256  & 0.099 & 0.009 & 0.607 & 0.19 & 1.41 & 15.25\\
432.6 & 113.3 & 0.0018  & 256  & 0.100 & 0.008 & 0.626 & 0.20 & 1.31 & 14.95\\
785.2 &  153.4 & 0.001   & 512  & 0.098 & 0.011 & 0.626 & 0.20 & 1.70 & 14.95\\
1026.3 &  176.9 & 0.00072 & 512  & 0.102 & 0.009 & 0.626 & 0.19 & 1.31 & 15.73\\
1529.0 &  217.0 & 0.0005  & 1024  & 0.100 & 0.008  & 0.63 & 0.19 & 2.02 & 18.80\\
2414.6 &  276.2 & 0.0003  & 1024 & 0.100 & 0.009 & 0.626 & 0.18 & 1.38 & 16.61\\
3535.0 &  335.2 & 0.0002  & 1024 & 0.102 & 0.008 & 0.626 & 0.18 & 1.01 & 16.61\\
5875.5 & 435.2 & 0.00011  & 2048 & 0.102 & 0.010  & 0.614 & 0.17 & 1.30 & 11.56
  \end{tabular}
 \end{center}
 \caption{A summary of the main parameters for our numerical simulations. 
 The values cited for the dissipation rate $\varepsilon$ and its standard 
 deviation $\sigma$, the rms velocity $U$, and the integral scale $L$,
 are ensemble- \emph{and} shell-averaged mean values. The quantity $t_{ss}/\tau$ denotes the time 
 the simulations have been run in steady state in units of large-eddy turnover time $\tau$. }
 \label{tbl:simulations}
\end{table}

Our simulations have been well validated by means of extensive and detailed
comparison with the results of other investigations \cite{Yoffe12, McComb14b}.
Furthermore, it can be seen from Fig.~\ref{fig:Ceps_Rl} that our results
reproduce the characteristic behavior for the plot of $\Ceps$ against $\Rl$,
and agree well with other representative results in the literature
\cite{Wang96,Cao99,Gotoh02,Kaneda03,Donzis05}. We note that the data
presented for comparison were obtained using negative damping (with variable
$\vep_W$) \cite{Kaneda03}, stochastic noise \cite{Gotoh02,Donzis05}, or
maintaining a $k^{-5/3}$ energy spectrum within the forced shells
\cite{Wang96,Cao99}. These methods for energy injection have been discussed
in Ref.~\cite{Bos07}.

\subsection{Results for dimensionless dissipation}

\begin{figure}[h!]
 \begin{center}
  \includegraphics[width=0.75\textwidth]{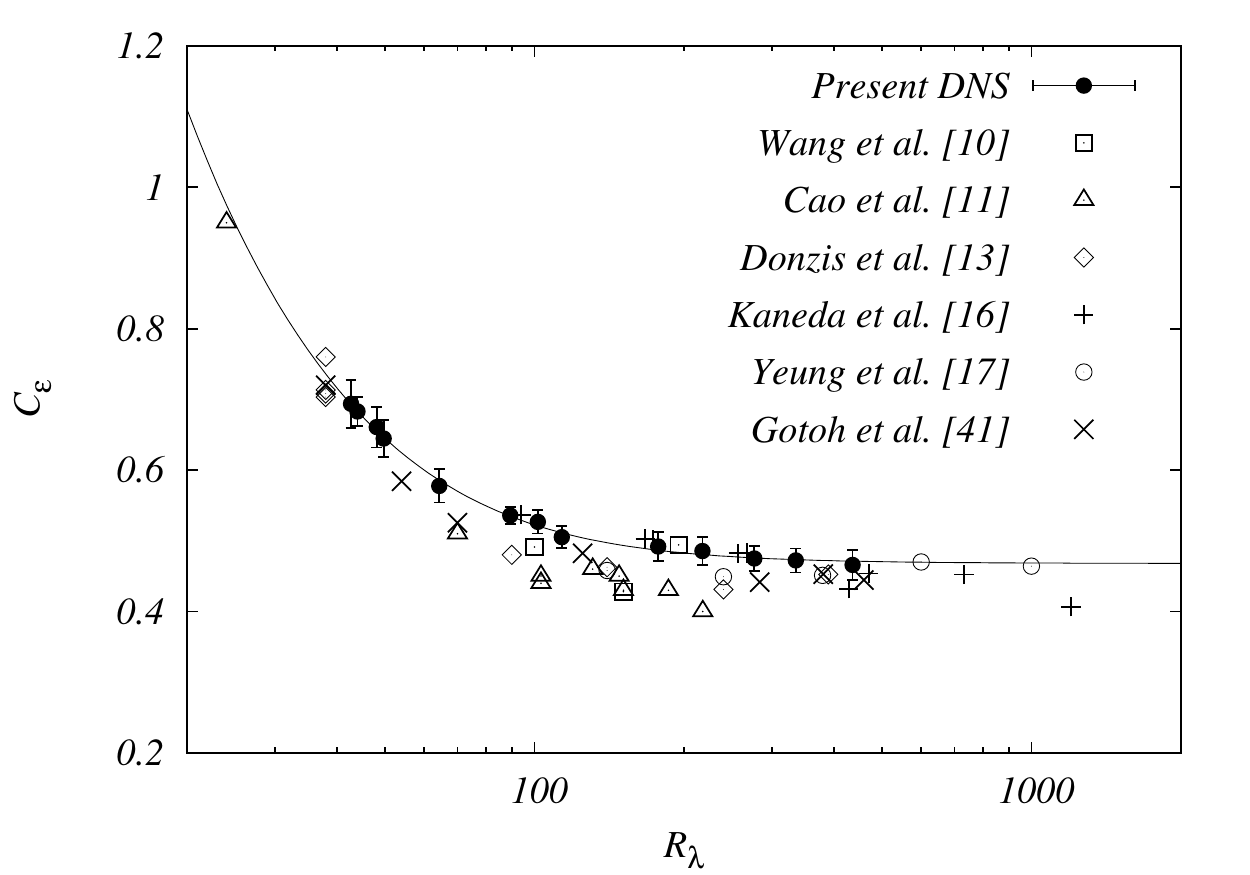}
 \end{center}
 \caption{ Variation of the dimensionless dissipation
coefficient $\Ceps$ with Taylor-Reynolds number $\Rl$ 
from our DNSs. Other investigations of forced
turbulence are presented for comparison. The black line is a 
fit of the expression \eqref{eq:Ceps_Rl} to our data only.}
 \label{fig:Ceps_Rl}
\end{figure}

Like other workers in the field, we follow the example of Sreenivasan in
plotting values of $\Ceps$ against $\Rl$ for various investigations.
Figure \ref{fig:Ceps_Rl} shows the values of $\Ceps$ obtained 
from our DNS alongside results from other investigations of forced isotropic
turbulence \cite{Wang96, Cao99, Gotoh02, Kaneda03, Donzis05, Yeung12}, plotted
against Taylor-Reynolds number. The black line is a fit of the expression 
\eqref{eq:Ceps_Rl}, which is equivalent
to the model equation \eqref{eq:DA_model}, to our data only, where the fit 
was carried out using the Marquardt-Levenberg least-squares method. The equivalence 
of the two expressions has been explained in Sec.~\ref{sec:asym_exp}.  
Recalling that \eqref{eq:Ceps_Rl} takes the form
\beq
\Ceps(\Rl) = A \big(1 + \sqrt{1 + (B/R_\lambda)^2} \big) \ ,
\eeq
we found the values $A= 0.234 \pm 0.003 $ and $B= 72 \pm 3$.

\begin{figure}[h!]
 \begin{center}
  \includegraphics[width=0.75\textwidth]{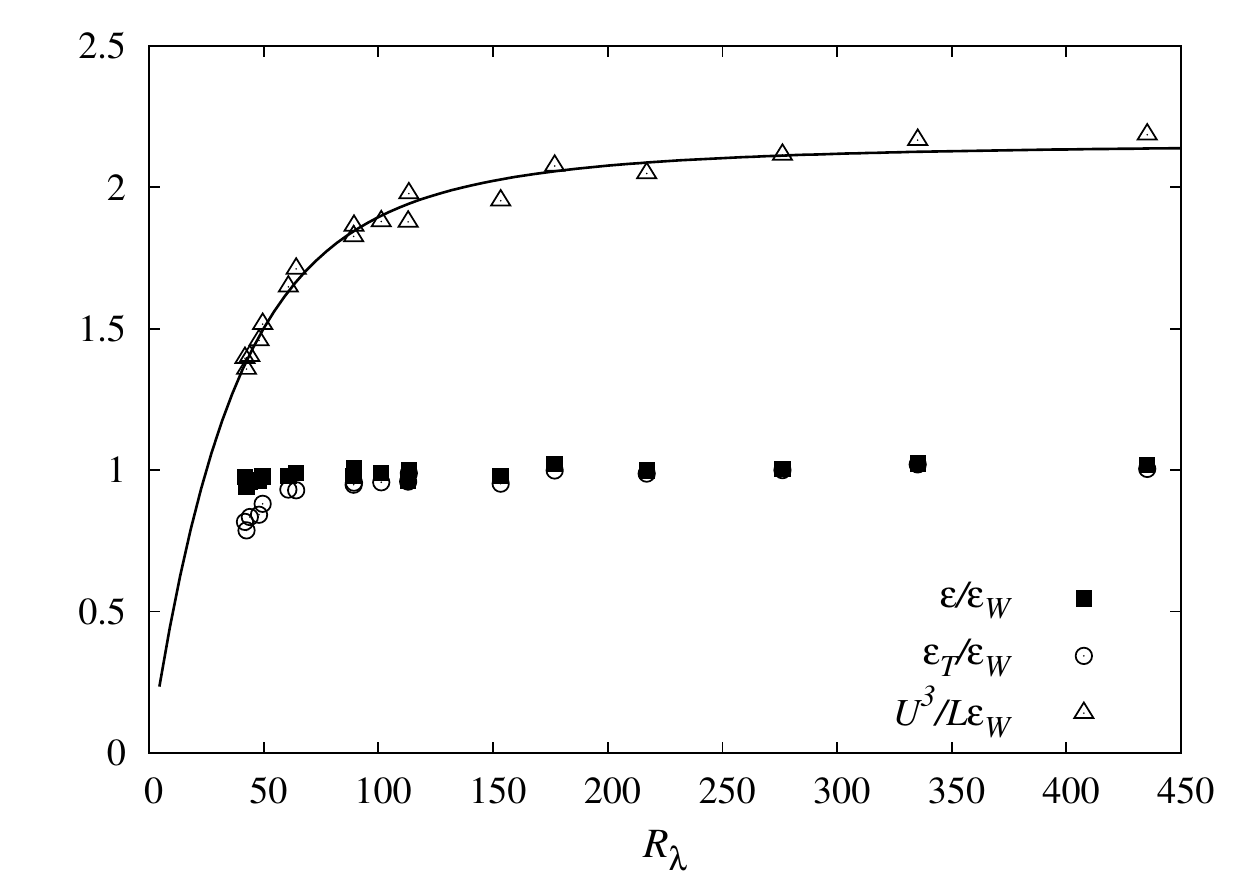}
 \end{center}
 \caption{ Variation with Taylor-Reynolds number of the
dissipation rate $\vep$, maximum inertial transfer rate $\vep_T$ and Taylor
surrogate $U^3/L$, all scaled on the injection rate $\vep_W$. The line 
corresponds to the fitted line in Fig.~\ref{fig:Ceps_Rl}.} 
 \label{fig:Taylor_surrogate}
\end{figure}

In Fig.~\ref{fig:Taylor_surrogate} we show separately the behavior of
the dissipation rate $\vep$, the maximum inertial flux $\vep_T$ and the
Taylor surrogate $U^3/L$, where each of these quantities was scaled on
the constant injection rate $\vep_W$.  We see that the decrease of
$\Ceps$, with increasing Reynolds number, is caused by the increasing
value of the surrogate in the denominator, rather than by decay of the
dissipation rate in the numerator, as this remains fixed at $\vep =
\vep_W$. This is the exact opposite of the case for freely decaying
turbulence, where the actual dissipation rate decreases with increasing
Reynolds number, while the surrogate remains fairly constant
\cite{McComb10b}. The figure also shows how both $\vep_T/\vep_W$ and
$U^3/(L\vep_W)$ increase at low $\Rl$, while $\vep/\vep_W$ is constant
(as required by the energy balance  in forced isotropic turbulence).
Therefore $U^3/L$ represents $\varepsilon_T$ better  than $\varepsilon$.
Furthermore, we observe that  $\varepsilon/\varepsilon_T =\vep_W/\vep_T
\to 1$ from above as the Reynolds number is increased, corresponding to
the onset of an inertial range \cite{McComb90a}.

\begin{figure}[!h]
 \begin{center}
  \includegraphics[width=0.75\textwidth]{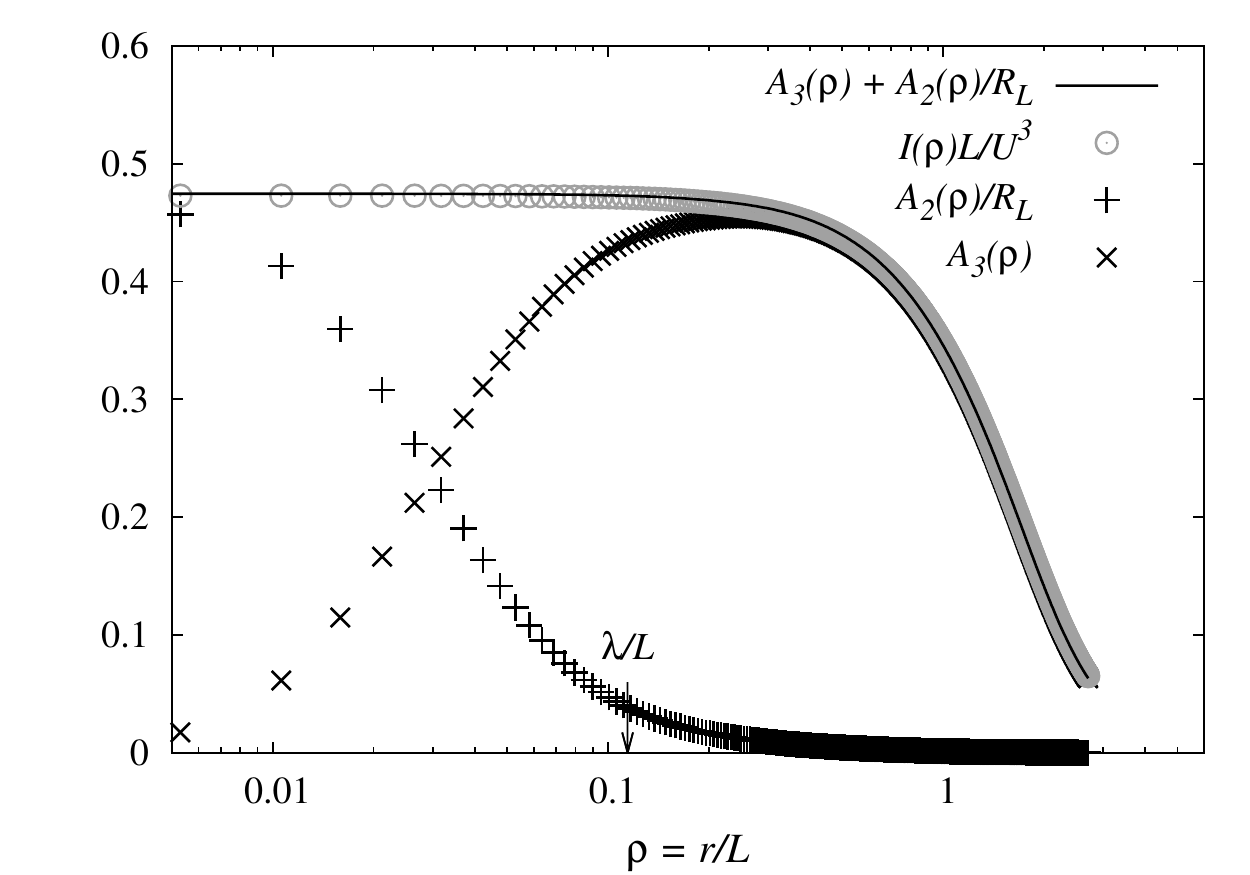}
 \end{center}
 \caption{ Dimensionless energy balance in the
KHE, as expressed by Eq.~\eqref{dim-input}.  
$R_\lambda = 276$. The Taylor microscale is labeled for comparison. Note that 
the energy input is constant for scales $r < \lambda$.}
 \label{fig:KHE_balance_dim}
\end{figure}

Figure \ref{fig:KHE_balance_dim} shows the balance of energy represented
by the dimensionless equation given as \eqref{dim-input}. For small scales 
($\rho < \lambda/L$ for the case $R_\lambda = 276$ shown) the input term 
satisfies $I(r) \simeq \varepsilon_W =\varepsilon$, as expected since such 
scales are not directly influenced by the
forcing. We note that the second- and third-order structure functions may be 
obtained from the energy and transfer spectra, respectively, using
\begin{equation}
\label{s2}
 S_2(r) = 4 \int_0^{\infty} dk\ E(k) \left( \frac{1}{3} - \frac{\sin{kr} - kr\cos{kr}}{(kr)^3} \right)  
\end{equation}
and
\begin{equation}
\label{s3}
 S_3(r) = 12 r \int_0^{\infty} dk\ T(k)  
\left( \frac{3\sin{kr} - 3kr\cos{kr}-(kr)^2\sin{kr}}{(kr)^5} \right) \ .
\end{equation}
This procedure
was introduced by Qian \cite{Qian97,Qian99} and more recently used by
Tchoufag \emph{et al} \cite{Tchoufag12} and by McComb \emph{et al}
\cite{McComb14b}: The underlying transforms may be found in the book by
Monin and Yaglom \cite{Monin75}; see their Eqs.~(12.75) and
(12.141$'''$). From these expressions, the nonlinear and viscous terms
$A_3$ and $A_2/R_L$  given by Eqs.~\eqref{eq:A3} and 
\eqref{eq:A2}, are calculated using
\beq
 A_3(\rho) = -\frac{3L}{U^3} \int_0^\infty dk\ T(k) \left[ 
 \frac{\sin{kL\rho} - kL\rho\cos{kL\rho}}{(kL\rho)^3} \right] \ ,
\eeq
an
\beq
\frac{A_2(\rho)}{\RL} = \frac{6\nu_0L}{U^3} \int_0^\infty dk\ k^2 E(k) \left[ 
 \frac{\sin{kL\rho} - kL\rho\cos{kL\rho}}{(kL\rho)^3} \right] .
\eeq

\begin{figure}[tbp]
 \begin{center} 
  \includegraphics[width=0.75\textwidth]{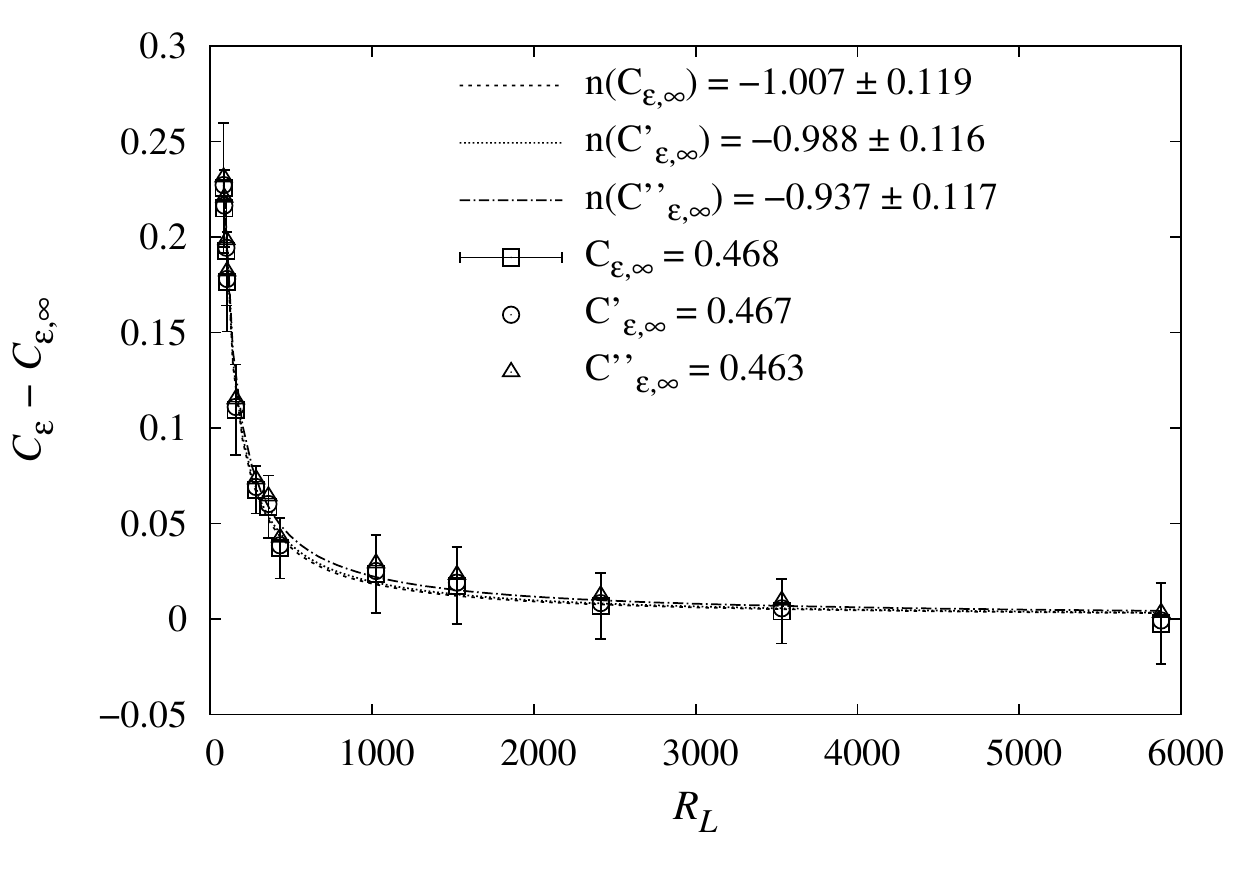}
 \end{center}
 \caption{
Graph of the present DNS results for 
 $\Ceps$ against Reynolds number, once the estimate of the 
asymptote is subtracted. 
The effect of varying our estimate of the asymptote $\Cinf$ is shown by the 
three different symbols, where $\Cinf^{\prime}$ and $\Cinf^{\prime \prime}$
denote variations in the asymptote within one standard error.
The dashed lines represent fits of the expression $CR_L^n$ to the data after 
subtracting the respective values of the asymptote.
}
 \label{fig:powerlaw_lin}
\end{figure}

\begin{figure}[tbp]
 \begin{center} 
  \includegraphics[width=0.75\textwidth]{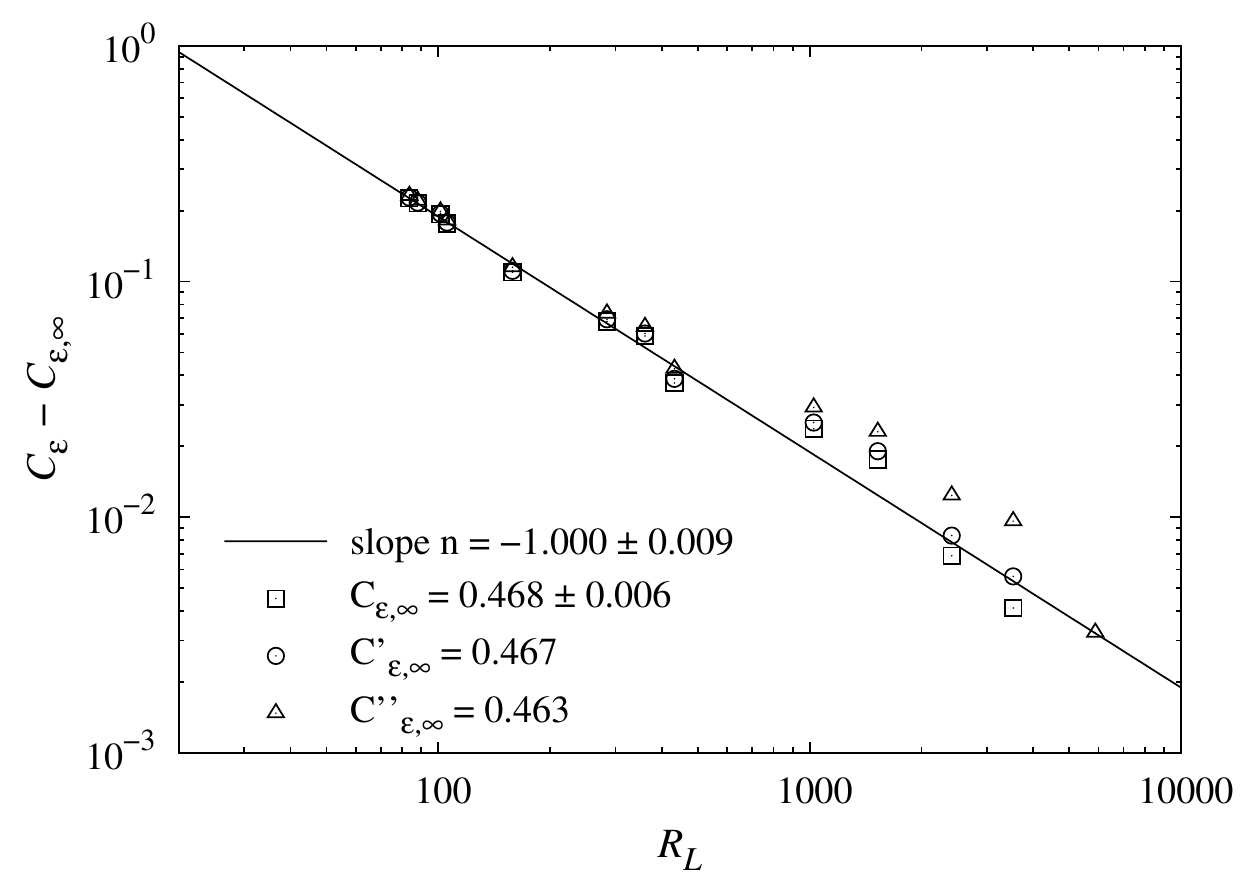}
 \end{center}
 \caption{
The same data as in Fig.~\ref{fig:powerlaw_lin} plotted on logarithmic scales. 
The solid line represents a slope of $n =-1.000 \pm 0.009$, obtained from a one-parameter
fit of the expression $C\RL^n$ to the data points, after subtracting the asymptote $\Cinf=0.468$.
}
 \label{fig:powerlaw_log}
\end{figure}

Figures \ref{fig:powerlaw_lin} and \ref{fig:powerlaw_log} show the measured power-law 
dependence of $\Ceps$ on $\RL$ on linear and logarithmic scales, respectively.
Noting that the standard procedure of using a log-log plot to 
identify power-law behavior is unavailable in this case, due to the 
constant asymptote, we subtracted the
estimated asymptotic value, which was obtained from a fit of \eqref{eq:DA_model} to DNS data
 (presented in the next section), and plotted $\Ceps - \Cinf$ against $\RL$ on
linear and logarithmic scales. This allowed us to identify power-law behavior 
consistent with $R_L^{-1}$. We also tested the effect of
varying our estimate of the value of the asymptote $\Cinf$. It can be
seen that the results were insensitive to this at the lower Reynolds 
numbers, where the $R_L^{-1}$-dependence is being tested. At higher $\RL$, the viscous 
contribution represented by $C/R_L$ becomes negligible and instead the result becomes
dependent on the actual value of $\Cinf$. 

As can be seen in Fig. \ref{fig:powerlaw_lin}, the value of the exponent $n$ depends 
weakly on the variation of the asymptote. The different values of $n$ shown in the figure 
were obtained by performing two-parameter fits of the expression $C\RL^n$ to the data points 
after subtracting the respective values of the asymptote. The fits using 
the asymptotes $\Cinf$, $\Cinf^{\prime}$, and $\Cinf^{\prime\prime}$, result in exponents consistent 
with a $1/R_L$ dependence of $\Ceps$ on $\RL$, 
namely, $n=-1.0 \pm 0.1$. The quality of the fit can be improved by fixing the 
coefficient $C$ to take the value $C=18.9$ obtained from the fit of \eqref{eq:DA_model}
to data, which is presented in the following section. 

\subsection{Assessment of the model}
\label{sec:model_review}

\begin{figure}[!tbp]
 \begin{center}
  \includegraphics[width=0.75\textwidth]{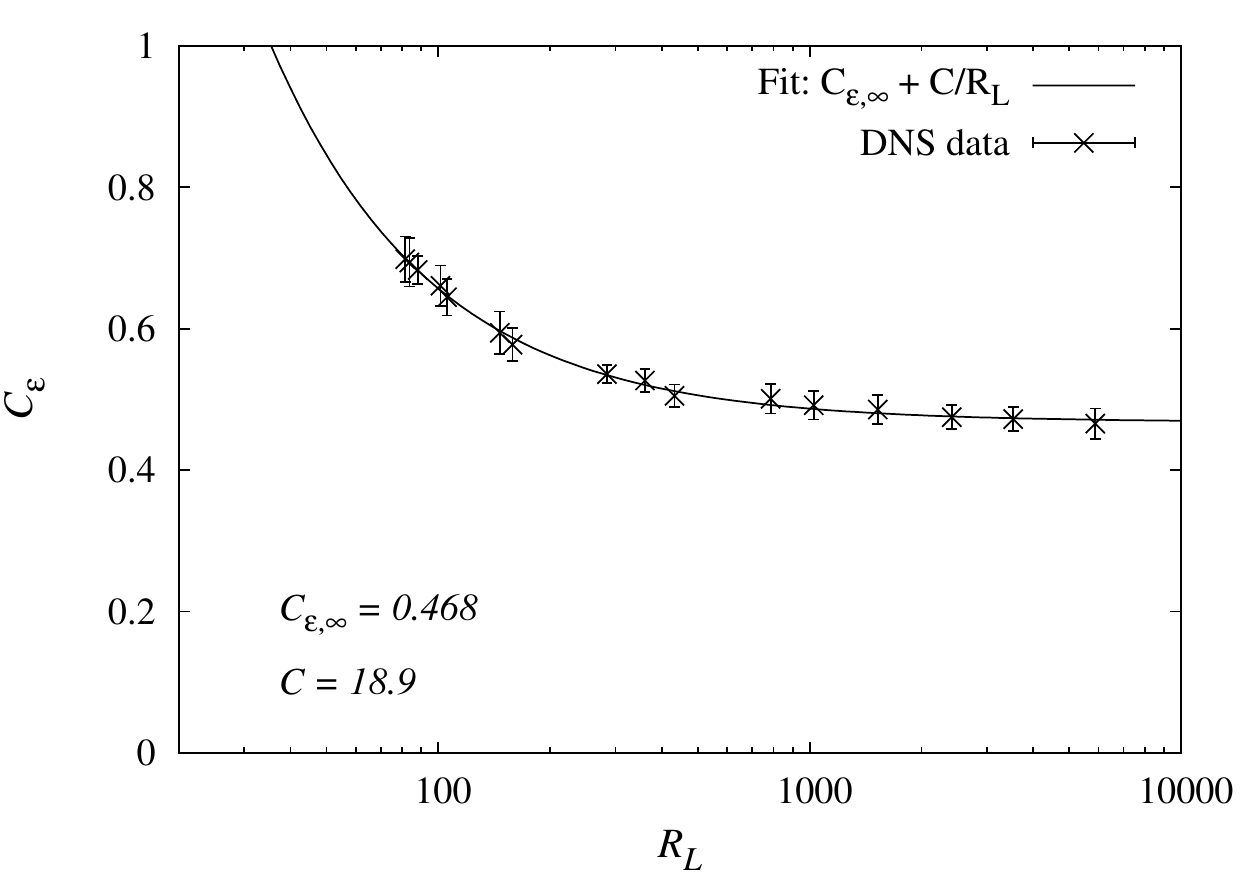}
 \end{center}
 \caption{The expression given in Eq.~\eqref{eq:DA_model} fitted 
 to present DNS data resulting in $\Cinf=0.468$ and $C=18.9$. 
}
 \label{fig:DA_model}
\end{figure}

In order to test our model for the dimensionless dissipation rate, we fitted 
an expression of the form
(\ref{eq:DA_model})
to data obtained with the present DNS, 
and it was found to agree very well, as shown 
in Fig.~\ref{fig:DA_model}. Measuring the 
exponent separately as explained in the previous section and shown 
in Fig~\ref{fig:powerlaw_lin}, resulted in  $n = -1.0 \pm 0.1$ and so supports the
model equation, with the constants given by $\Cinf = 0.468 \pm 0.006$ and $C 
=18.9 \pm 1.3$. 
Fixing the value of the coefficient $C$ to be $C=18.9$, as obtained by 
the fit of \eqref{eq:DA_model} to data, and by performing a one-parameter fit, 
varying only the exponent, results in $n=-1.000\pm 0.009$, as shown in 
Fig.~\ref{fig:powerlaw_log}.  

As shown in Fig.~\ref{fig:DA_model} (and in Fig.~\ref{fig:Ceps_Rl}),
it may be seen that our model \eqref{eq:DA_model} is in good agreement with 
both our own data and that of others, where we note that the expression fitted to 
our data in Fig.~\ref{fig:Ceps_Rl} is equivalent to our model \eqref{eq:DA_model}.

\section{Discussion}

Our model, as given by either Eq.~\eqref{eq:DA_model} (for
dependence on $\RL$) or Eq.~\eqref{eq:Ceps_Rl} (for dependence on
$\Rl$), may be compared to other work in the literature. As mentioned in
the  Introduction, Sreenivasan  \cite{Sreenivasan84} compared
experimental results for free decay to the  expression for very low
Reynolds numbers,
\begin{equation} \label{eq:C_sreeni}
 \Ceps = \frac{15}{R_\lambda} \sqrt{\frac{\pi}{2}} \ .
\end{equation}
This used the isotropic relation $\varepsilon = 15\nu_0 U^2/\lambda^2$ 
(where $\lambda$ is the Taylor microscale) and the approximation $L/\lambda 
\simeq (\pi/2)^{1/2} \ $ \cite{Batchelor71}. Note that, while $15\sqrt{\pi/2} = 
18.8$, compared to $C = 18.9 \pm 1.3$ found in the present analysis, this 
expression involves $R_\lambda$ rather than $R_L$.
At low $R_L$, however, $R_L \sim \Rl$; thus, by combination of the two 
asymptotic results in Sreenivasan's paper \cite{Sreenivasan84}, one obtains 
the result for the scaling of the dimensionless dissipation rate reported here. 
Furthermore, the values of the coefficient $C$ obtained by Sreenivasan
and measured numerically by us agree within one 
standard error.  

Later, Lohse \cite{Lohse94} used ``variable range mean-field theory" to find 
an expression for the dimensionless dissipation coefficient by matching 
small $r$ and inertial range forms for the second-order structure function, 
and obtained
\begin{align}
 \Ceps 
  &= \Cinf \sqrt{1 + \frac{5b^3}{4R_\lambda^2}} \ ,
\end{align}
where $b = S_2(r)/(\vep r)^{2/3}$ such that $\Cinf = (h_2(1)/b)^{3/2}$. At 
low Reynolds numbers, this author reported $\Ceps = 18/R_L$. The asymptotic 
value was calculated by Pearson, Krogstad and van der Water 
\cite{Pearson02}, who used $h_2(1) \simeq 1.25$ and $b \simeq 2.05$,
to be $\Cinf \simeq 0.48$, which agrees with our result, $\Cinf = 0.468 \pm 0.006$,
 nearly within one standard error. 

In an alternative approach, Doering and Foias \cite{Doering02} used the 
longest lengthscale affected by forcing, $l$, to derive upper and lower 
bounds on $\Ceps$,
\begin{align}
 \label{eq:DF_ineq}
 \frac{4\pi^2}{\alpha^2 Re} \leq \Ceps \leq \left( \frac{a}{Re} + b \right) 
 \end{align}
for constants $a,b$, where $Re = Ul/\nu_0$ and $\alpha = L_\text{box}/l$.
While the upper bound resembles the present model, it is important to note 
that where these authors have obtained an inequality, we have an equality. 
Inspired by the results of \cite{Doering02}, Eq. \eqref{eq:Ceps_Rl}, which is 
equivalent to the model equation \eqref{eq:DA_model} and thus to the expression 
in the upper bound \eqref{eq:DF_ineq}, was 
fitted to data by Donzis, Sreenivasan and Yeung \cite{Donzis05}, 
with $A \simeq 0.2$ and $B \simeq 92$ giving reasonable agreement, such 
that $\Cinf \simeq 0.4$.

Later still, Bos, Shao and Bertoglio \cite{Bos07} employed the idea of a
 finite cascade time to relate the expressions for $\Ceps$ in forced and
 decaying turbulence. Using a model spectrum, they then derived a form
for  $\Ceps$ and found the asymptotic value $\Cinf = 0.53$ with the
Kolmogorov  constant $C_K = 1.5$. Note that when we used their formula,
with the value $C_K = 1.625$ instead (which is probably more representative 
\cite{McComb14a}), this led to $\Cinf = 0.47$, as
found in the present work. With a simplified model spectrum, the 
authors then showed how their expression reduced to $\Ceps = 19/R_L$ for
low  Reynolds numbers [when $E(k) \sim k^4$ at low $k$] in agreement with
$C =  18.9\pm 1.3$ found here (within one standard error).

We finish with a brief consideration of the universality of these
results. In general this would mean that the $\Ceps$ versus $\Rl$ curve
would take the same form for all flow configurations, such as pipe flow,
free jets, isotropic turbulence, and so on. Evidently, as our present
work is restricted to stationary isotropic turbulence, this rather
restricts what we can say about the matter. Indeed, we basically can
only consider the effects of the initial conditions such as the form of
the forcing and the shape of the initial spectrum, and insofar as these
have been tested, our brief literature survey would indicate that they
probably only affect the duration of transient behavior, but not the
steady-state results. This is, of course, in line with  what one expects
from universality of isotropic turbulence in general. That is, forcing
should be confined to low wavenumbers in order to set up an asymptotic
state which is representative of the equations of motion, rather than
the arbitrarily chosen forcing. Similarly, the arbitrary initial energy
spectrum should quickly die away to be replaced by the true spectrum. So
it is important to recognize that the universality of the $\Ceps$ curve
should be considered in conjunction with the universality of the
turbulence that we are producing. Our present work suggests that the
model based on $\delta$-function forcing is in good agreement with the DNSs
based on finite (in wavenumber space) forcing, and that the values of
the constants $C$ and $\Cinf$ agree quite well with those obtained in
other investigations. This might be seen as evidence for universality
within the confines of this particular flow. Certainly one should
observe that the scatter of points from various investigations in 
Fig.~\ref{fig:Ceps_Rl} is \emph{not} evidence of nonuniversality, 
unless one has eliminated other possible explanations for this scatter, 
such as differences in run time or resolution. 

We made a systematic investigation into the effect of run time
on the measured value of $\Ceps$ for our highest $R_L$ data point.
In total, this run was carried out for about 12 large-eddy turnover times in steady state,
resulting in the measured value of $\Ceps = 0.466 \pm 0.021$. If we restrict the
time interval that we average results over to, say, 3 large-eddy turnover times,
we measure $\Ceps =0.442 \pm 0.030$, which is significantly lower than the measured
value averaged over the full run. Note that the value obtained from the shorter
time interval is closer to some of the values measured by other groups shown in Fig.~\ref{fig:Ceps_Rl}.

Then, by extending the time interval systematically towards the actual run time in steady
state, we found that the results converged to the value obtained by averaging
over the full time interval. Work on this aspect continues as part of our program 
on DNS and will be reported in due course.

\section{Conclusions}

Our theoretical model predicts an inverse dependence of the
dimensionless dissipation rate on the integral scale Reynolds number,
with asymptotic validity in the limit of large Reynolds numbers. A
question then arises: Do we have to include higher-order terms at lower
Reynolds numbers? It is in order to answer this question that we resort
to direct numerical simulation.

The answer to our question is reassuring. We find that analysis of the
data from our DNS supports a dependence on $\RL^{-1}$ at all values of
the Reynolds number. Also, the law given by Eq.~(\ref{eq:DA_model})
is found to give a good fit to the data from the DNS, with values for the
constants which are in generally good agreement with those obtained in
other investigations.

It may be of interest to note, that when we apply the same theoretical
approach to magnetohydrodynamics (MHD), we find that it is necessary to 
take the term in $R_L^{-2}$ into account, in addition to the leading order term, although
the effect was not large \cite{LinkmannETCb}. We also plan to extend the 
analysis to inhomogeneous flows, in order to examine further the question of
universality, as discussed at the end of the preceding section.

Last, we note that our analysis shows that the behavior of the
dimensionless dissipation rate, as found experimentally, is entirely in
accord with the Kolmogorov (K41) picture of turbulence and, in
particular, with Kolmogorov's derivation of his four-fifths law
\cite{Kolmogorov41b}, the one universally accepted result in turbulence.

\acknowledgments
This work has made use of the resources provided by HECToR 
({\tt{http://www.hector.ac.uk/}}), made available through 
ECDF ({\tt{http://www.ecdf.ed.ac.uk/}}). 
A.~B. is supported by STFC, S.~R.~Y. and M.~F.~L. are funded by EPSRC. 

\bibliography{wdm}

\end{document}